\documentclass[epj,nopacs]{svjour}
\usepackage{bm}
\usepackage{amsmath}

\usepackage{widetext}
\usepackage{cite}
\usepackage[bbgreekl]{mathbbol}
\usepackage{mhchem}
\usepackage{slashed}
\usepackage{subfigure}
\usepackage{amssymb}
\usepackage{mathrsfs}
\usepackage{hyperref}
\usepackage{graphics}
\def\p{\mbox{\boldmath$\displaystyle\mathbf{p}$}}

\def\bv{\mbox{\boldmath$\displaystyle\mathbf{\varphi}$}}

\def\0{\mbox{\boldmath$\displaystyle\mathbf{0}$}}
\def\s{\mbox{\boldmath$\displaystyle\mathbf{\sigma}$}}

\def\x{\mbox{\boldmath$\displaystyle\mathbf{x}$}}
\def\y{\mbox{\boldmath$\displaystyle\mathbf{y}$}}
\begin{document}

\title{Fermionic degeneracy and non-local contributions in flag-dipole spinors and mass dimension one fermions}
\author{Cheng-Yang Lee
\thanks{email: cylee@scu.edu.cn}%
}                     
%
%
\institute{Manipal Centre for Natural Sciences, Centre of Excellence,\\ Manipal Academy of Higher Education\\ Dr. T. M. A. Pai Planetarium Building, Manipal 576104, Karnataka, India\\
\\
Center for Theoretical Physics, College of Physical Science and Technology,
\\Sichuan University, Chengdu, 610064, China
}
\date{Received: date / Revised version: date}
%
\abstract{
We construct a mass dimension one fermionic field associated with flag-dipole spinors. These spinors are related to Elko (flag-pole spinors) by a one-parameter matrix transformation $\mathcal{Z}(z)$ where $z$ is a complex number. The theory is non-local and non-covariant. While it is possible to obtain a Lorentz-invariant theory via $\tau$-deformation, we choose to study the effects of non-locality and non-covariance. Our motivation for doing so is explained. We show that a fermionic field with $|z|\neq1$ and $|z|=1$ are physically equivalent. But for fermionic fields with more than one value of $z$, their interactions  are $z$-dependent thus introducing an additional fermionic degeneracy that is absent in the Lorentz-invariant theory. We study the fermionic self-interaction and the local $U(1)$ interaction. In the process, we obtained non-local contributions for fermionic self-interaction that have previously been neglected. For the local $U(1)$ theory, the interactions contain time derivatives that renders the interacting density non-commutative at space-like separation. We show that this problem can be resolved by working in the temporal gauge. This issue is also discussed in the context of gravity.
\PACS{
      {12.60.-i}{Models beyond the Standard Model}
     } 
} 

\maketitle
%
\section{Introduction}

In the Standard Model (SM), the Dirac and Weyl spinors and their quantum fields have both played important roles in describing the dynamics of the fermions as well as elucidating the structure of the model. Despite its success, there remains many outstanding questions which led to the common consensus that the model is incomplete. 

In this paper, we propose a promising approach to investigate the physics beyond the SM. This is based on the fact that according to the Lounesto classification, the Dirac and Weyl spinors are not the only spin-half representations of the Lorentz group. Instead, there exists six classes of spinors, each uniquely defined by their bilinear covariants~\cite[sec.~12]{Lounesto:2001zz}. Let $\psi$ be an arbitrary spinor that transforms under the $(\frac{1}{2},0)\oplus(0,\frac{1}{2})$ representation of the Lorentz group. The bilinear covariants needed for the classification are (the vector current $J^{\mu}=\overline{\psi}\gamma^{\mu}\psi$ is also included but it is not directly relevant in the classification):
\begin{equation}
\begin{array}{ll}
\Omega_{1}=\overline{\psi}\psi, &\Omega_{2}=\overline{\psi}\gamma^{5}\psi,\\
K^{\mu}=\overline{\psi}\gamma^{5}\gamma^{\mu}\psi,&
 S^{\mu\nu}=\overline{\psi}\gamma^{\mu}\gamma^{\nu}\psi
\end{array}
\end{equation}
where $\overline{\psi}=\psi^{\dag}\gamma^{0}$ is the Dirac dual and the $\gamma^{\mu}$ matrices are chosen to be
\begin{equation}
\begin{array}{lll}
\gamma^{0}=
\left(\begin{matrix}
O & I \\
I & O
\end{matrix}\right),&
\gamma^{i}=
\left(\begin{matrix}
O & -\sigma^{i} \\
\sigma^{i} & O
\end{matrix} \right),&
\gamma^{5}=
\left(\begin{matrix}
I & O \\
O &-I \end{matrix}
\right).
\end{array} \label{eq:gamma}
\end{equation}
The Lounesto classification is given by
\begin{enumerate}
\item $\Omega_{1}\neq0$, $\Omega_{2}\neq 0$
\item $\Omega_{1}\neq0$, $\Omega_{2}=0$
\item $\Omega_{1}=0, \Omega_{2}\neq 0$
\item $\Omega_{1}=0,$ $\Omega_{2}=0$, $K^{\mu}\neq0$, $S^{\mu\nu}\neq0$
\item $\Omega_{1}=0,$ $\Omega_{2}=0$, $K^{\mu}= 0$, $S^{\mu\nu}\neq0 $
\item $\Omega_{1}=0,$ $\Omega_{2}=0$, $K^{\mu}\neq 0$, $S^{\mu\nu}=0$
\end{enumerate}

In the chosen basis given by eq.~(\ref{eq:gamma}), the Dirac spinors that are associated with the fermionic field belong to the 2nd class. In the original work, Lounesto identified the 1st-3rd classes to be the Dirac spinors. It is unclear to us why this identification was made but since this is not the focus of this paper we will not pursue it further here. The 4th and 5th classes are known as the \textit{flag-dipole} and \textit{flag-pole spinors} respectively. The 6th class is the Weyl spinors.

From the Lounesto classification, a natural question arises --\textit{What are the quantum field theories associated with the flag-dipole and flag-pole spinors?} This question was partially answered in a series of publications on the theory of mass dimension one fermionic field with Elko (flag-pole spinors) as expansion coefficients~\cite{Ahluwalia:2004sz,Ahluwalia:2004ab,daRocha:2005ti,
daRocha:2007pz,daRocha:2009gb,daRocha:2008we,HoffdaSilva:2009is,Ahluwalia:2008xi,Ahluwalia:2009rh,
Fabbri:2009ka,Fabbri:2009aj,Fabbri:2010va,daRocha:2011yr,daRocha:2011xb,Lee:2012td,Lee:2014opa,Lee:2015jpa,Lee:2015sqj,Ahluwalia:2016jwz,Ahluwalia:2016rwl,Cavalcanti:2014uta,Nikitin:2014fga,Ahluwalia:2019etz}. These fermionic fields have two important properties. They are of mass dimension one and satisfy the Klein-Gordon but not the Dirac equation. The mass dimension one fermions have been studied as a dark matter candidate. Their signatures at the Large Hadron Collider  and in cosmology have been investigated~\cite{Dias:2010aa,Alves:2014kta,Agarwal:2014oaa}. Their gravitational interactions have also received much attention~\cite{Boehmer:2006qq,Boehmer:2007dh,Boehmer:2008rz,Boehmer:2007ut,Boehmer:2008ah,Boehmer:2009aw,Shankaranarayanan:2009sz,Boehmer:2010ma,
Gredat:2008qf,Wei:2010ad,Basak:2012sn,daRocha:2013qhu,daSilva:2014kfa,daRocha:2014dla,Rogerio:2019evl}. 

With the discovery of Elko and mass dimension one fermions, to answer the above question, the remaining task is to study the flag-dipole spinors. Using results obtained in~\cite{Cavalcanti:2014wia,Ahluwalia:2016rwl}, we construct a mass dimension one fermionic field with flag-dipole spinors. The flag-dipole spinors are shown to be related to Elko by a one-parameter matrix transformation $\mathcal{Z}(z)$ where $z$ is a non-zero complex number. 


The spin-sums for Elko and flag-dipole spinors are not Lorentz-covariant. As a result, their quantum field theories are non-local, non-covariant and is endowed with a preferred direction. In what is to follow, for brevity, we shall simply refer to the theory as being non-covariant. By introducing the $\tau$-deformation to the spinor dual~\cite{Ahluwalia:2016rwl}, one can show that both theories are Lorentz-invariant, physically equivalent and $z$-independent~\cite{Rogerio:2019xcu}. However, this is not the approach we wish to take. Instead, we will study the non-covariant theory. One is justified to ask -- \textit{Why study a non-covariant theory?} There are several reasons. 
The Lorentz-invariant formulation is obtained by removing a Lorentz-violating $\mathcal{G}$-matrix from the spin-sums. While $\mathcal{G}$ is non-covariant, it is important to note that it naturally appears from the Elko construct and has natural generalizations to higher-spin~\cite{Lee:2019fni}. Since the non-locality and non-covariance induced by $\mathcal{G}$ is not postulated, it is important to understand these properties. In our opinion, the removal of $\mathcal{G}$ by $\tau$-deformation is likely to be correct and well-justified since it yields a Lorentz-invariant theory. However, there remains issues to be resolved. To implement the deformation, the involved matrix multiplication is non-associative~\cite{Lee:2019fni}. This is an unusual feature so further investigation is needed to verify its legitimacy. 

In the non-covariant theory, the spin-sums and propagator are $z$-dependent. We find, for fermionic fields with only one $|z|\neq1$, it is physically equivalent to its Elko counterpart with $|z|=1$. But for fermionic fields with more than one value of $z$, their interactions become $z$-dependent. This suggests that the fermions have an additional degeneracy specified by $z$, a feature that is absent in the Lorentz-invariant theory.

In the existing literature, the effects of non-covariance that originates from the spin-sums and propagator have been investigated~\cite{Agarwal:2014oaa,Lee:2015sqj,Lee:2015jpa}. However, the effects of non-local anti-commutators have not been taken into account. That is, one simply write down the local interacting potential in the interacting picture and assume it to be the only contribution to the $S$-matrix. This procedure turns out to be incorrect. The non-local anti-commutators do affect the interactions. To show this, we study the fermionic self-interaction and the local $U(1)$ interaction. We derive the field equation in the Heisenberg picture using the prescription given in ref.~\cite{Weinberg:1965rz}. We find, there are indeed non-local contributions that have been missed in the previous works. The non-local contributions do not seem to cause difficulties for the theory. For the local $U(1)$ interaction, the interacting Hamiltonian density is in general non-commutative at space-like separation which leads to causality violation of the $S$-matrix. This is a generic feature for mass dimension one fermions with a preferred direction whose interactions involve time derivatives. We show that the problem can be resolved by working in the temporal gauge where the time component of the gauge field vanishes thus preserving causality of the $S$-matrix. For interactions with gravity, a similar issue also arise and will be discussed.



The paper is organized as follows. In sec.~\ref{fp}, we construct the flag-dipole spinors and show that they are related to Elko by a matrix transformation. In sec.~\ref{ncf} and app.~\ref{self}, we study the non-covariant theory.


\begin{widetext}
\section{Flag-dipole spinors} \label{fp}

In ref.~\cite{Cavalcanti:2014wia}, by explicitly computing the bilinear covariants, Calvalcanti showed that the most general flag-dipole spinors are given by
\begin{eqnarray}
&&\left(\begin{matrix}
\mbox{Flag-dipole spinors with}\\
\mbox{two non-zero components, $\Psi_{2}$}
\end{matrix}\right)=
\left\{\left(\begin{matrix}
a_{1} \\
0 \\
0 \\
a_{2}
\end{matrix}\right), \quad
\left(\begin{matrix}
0 \\
a_{3} \\
a_{4} \\
0
\end{matrix}\right), \quad
|a_{1}|^{2}\neq|a_{2}|^{2}, 
|a_{3}|^{2}\neq|a_{4}|^{2}\right\} \label{eq:dipole2}\\
&&\left(\begin{matrix}
\mbox{Flag-dipole spinors with}\\
\mbox{four non-zero components, $\Psi_{4}$}
\end{matrix}\right)=
\left\{\left(\begin{matrix}
-a_{5}a_{6}a_{7}^{*}/|a_{6}|^{2} \\
a_{5}\\
a_{6}\\
a_{7}\\
\end{matrix}\right), \quad
|a_{5}|^{2}\neq |a_{6}|^{2}\right\}
\label{eq:dipole4}
\end{eqnarray}
where $a_{1},\cdots,a_{7}$ are arbitrary complex numbers. To understand their physics, we note that while these solutions are linearly independent, $\Psi_{2}$ and $\Psi_{4}$ are related by a Lorentz boost
\begin{eqnarray}
L&=&\left[\begin{matrix}
\exp\left(\frac{1}{2}\s\cdot\bv\right) & O \\
O & \exp\left(-\frac{1}{2}\s\cdot\bv\right)
\end{matrix}\right]
\end{eqnarray}
where $\s=(\sigma_{x},\sigma_{y},\sigma_{z})$ are the Pauli matrices and $\bv=\varphi\hat{\p}$ is the rapidity parameter with $\cosh\varphi=E/m$, $\sinh\varphi=|\p|/m$. Specifically, taking $\psi_{2}\in\Psi_{2}$ we find that $L\psi_{2}\in\Psi_{4}$. In light of this observation, it is only necessary to consider flag-dipole spinors belonging to $\Psi_{2}$.
\end{widetext}

The above solutions for the flag-dipole spinors do not necessarily guarantee the existence of a physical theory. By  physical, we mean that the associated fermionic fields must have positive-definite free Hamiltonian and preserve causality. After imposing these conditions, we find that the flag-dipole spinors in the helicity basis which give rise to a physical fermionic field, is related to Elko given in ref.~\cite{Ahluwalia:2009rh} by a matrix transformation\footnote{In this paper, the Elko solutions are in the helicity basis. For Elko in the polarization basis, a subtlety arises. Please see eqs.~(17) and (18) in ref.~\cite{Ahluwalia:2009rh} for details.}
\begin{eqnarray}
&&\lambda^{S}_{\sigma,z}(k^{\mu})=\mathcal{Z}(z)\lambda^{S}_{\sigma}(k^{\mu}),\label{eq:S}\\
&&\lambda^{A}_{\sigma,z}(k^{\mu})=\mathcal{Z}(z)\lambda^{A}_{\sigma}(k^{\mu})\label{eq:A}
\end{eqnarray}
where $\mathcal{Z}(z)$ is a matrix given by
\begin{equation}
\mathcal{Z}(z)=\left(\begin{matrix}
{z^{*}}^{-1} I & O\\
O & z I \end{matrix}\right)
\end{equation}
and $z$ is a complex number. For $|z|^{2}\neq1$, we obtain the flag-dipole spinors. Since $\mathcal{Z}$ is comprised of the identity matrix, it commutes with all the Lorentz generators of the $(\frac{1}{2},0)\oplus(0,\frac{1}{2})$ representation. Therefore, eqs.~(\ref{eq:S}) and (\ref{eq:A}) are valid for all momentum
\begin{eqnarray}
&&\lambda^{S}_{\sigma,z}(p^{\mu})=\mathcal{Z}(z)\lambda^{S}_{\sigma}(p^{\mu}),\\
&&\lambda^{A}_{\sigma,z}(p^{\mu})=\mathcal{Z}(z)\lambda^{A}_{\sigma}(p^{\mu}).\label{eq:dipole3}
\end{eqnarray}
Consequently, we obtain
\begin{align}
\mathfrak{f}_{z}(x)=\mathcal{Z}(z)\mathfrak{f}(x)
\end{align}
where $\mathfrak{f}(x)$ is the mass dimension one fermionic field given in ref.~\cite{Ahluwalia:2016rwl}. The matrix $\mathcal{Z}$ is dimensionless so $\mathfrak{f}_{z}$ is still of mass dimension one. The dual for the flag-dipole spinors are related to their Elko counterparts by
\begin{align}
&\widetilde{\lambda}^{S}_{\sigma,z}(p^{\mu})=\widetilde{\lambda}^{S}_{\sigma}(p^{\mu})\mathcal{Z}^{-1}(z),\\
&\widetilde{\lambda}^{A}_{\sigma,z}(p^{\mu})=\widetilde{\lambda}^{A}_{\sigma}(p^{\mu})\mathcal{Z}^{-1}(z)
\end{align}
where
\begin{align}
&\widetilde{\lambda}^{S}_{\sigma}(p^{\mu})\equiv i(-1)^{1/2+\sigma}\overline{\lambda}^{S}_{-\sigma}(p^{\mu}),\\
&\widetilde{\lambda}^{A}_{\sigma}(p^{\mu})\equiv i(-1)^{1/2+\sigma}\overline{\lambda}^{A}_{-\sigma}(p^{\mu}). \label{eq:edual1}
\end{align}
Consequently, we have
\begin{equation}
\widetilde{\mathfrak{f}}_{z}(x)=\widetilde{\mathfrak{f}}(x)\mathcal{Z}^{-1}(z).
\end{equation}

Taking $z=|z|e^{i\alpha}$, we note that $\alpha$ is a global phase associated with $\mathcal{Z}(z)$ so its effect is unimportant. Therefore, we may simply take $\mathcal{Z}$ to be
\begin{equation}
\mathcal{Z}(z)=\left(\begin{matrix}
|z|^{-1}I & O \\
O & |z|I
\end{matrix}\right).
\end{equation}
The spin-sums for the flag-dipole spinors are given by
\begin{eqnarray}
&& \sum_{\sigma}\lambda^{S}_{\sigma,z}(p^{\mu})\widetilde{\lambda}^{S}_{\sigma,z}(p^{\mu})=
+m[I+\mathcal{G}_{z}(\p)], \\
&& \sum_{\sigma}\lambda^{A}_{\sigma,z}(p^{\mu})\widetilde{\lambda}^{A}_{\sigma,z}(p^{\mu})=
-m[I-\mathcal{G}_{z}(\p)] \label{eq:spin_sums}
\end{eqnarray}
where $\mathcal{G}_{z}(\p)$ is defined as
\begin{align}
\mathcal{G}_{z}(\p)=
\left(\begin{matrix}
0 & 0 & 0 & -ie^{-i\phi}|z|^{-2}\\
0 & 0 & ie^{i\phi}|z|^{-2} & 0 \\
0 & -ie^{-i\phi}|z|^{2} & 0 & 0 \\
ie^{i\phi}|z|^{2} & 0 & 0 & 0
\end{matrix}\right)
\end{align}
with $0\leq\phi\leq2\pi$ being the azimuthal angle in the $xy$-plane. The non-covariance of the spin-sums implies that the theory is Lorentz-violating. The non-covariance can be removed by defining a new dual with an infinitesimal deformation~\cite{Ahluwalia:2016rwl}. As a result, the fermionic fields constructed from the flag-dipole spinors and Elko are physically equivalent~\cite{Rogerio:2019xcu}.

\section{The non-covariant formulation} \label{ncf}

While there is a Lorentz-invariant formulation, in our opinion, there are still good reasons to study the non-covariant theory. As evident from discussions in~\cite{Ahluwalia:2008xi,Ahluwalia:2009rh}, the effects of the $\mathcal{G}$ matrix is not redundant and it also admits natural higher-spin generalizations~\cite{Lee:2019fni}. Therefore, to understand the theory in its totality, it is important to study the effects of non-covariance and non-locality due to $\mathcal{G}$. Even if this turns out to be non-physical, we should still be informed for the sake of completeness. 

From the spin-sums given in eq.~(\ref{eq:spin_sums}), the Lagrangian and propagator are given by
\begin{eqnarray}
&& \mathscr{L}(x)=\partial^{\mu}\widetilde{\mathfrak{f}}_{z}(x)\partial_{\mu}\mathfrak{f}_{z}(x)-m^{2}\widetilde{\mathfrak{f}}_{z}(x)\mathfrak{f}_{z}(x),\label{eq:L}\\
&& S_{z}(p)=\frac{i}{(2\pi)^{4}}\frac{I+\mathcal{G}_{z}(\p)}{p^{2}-m^{2}+i\epsilon} \label{eq:prop}.
\end{eqnarray}
The flag-dipole spinors remains orthonormal under the dual
\begin{align}
&\widetilde{\lambda}^{S}_{\sigma,z}(p^{\mu})\lambda^{S}_{\sigma',z}(p^{\mu})=-
\widetilde{\lambda}^{A}_{\sigma,z}(p^{\mu})\lambda^{A}_{\sigma',z}(p^{\mu})=2m\delta_{\sigma\sigma'},\\
&\widetilde{\lambda}^{S}_{\sigma,z}(p^{\mu})\lambda^{A}_{\sigma',z}(p^{\mu})=0.
\end{align}
so the Hamiltonian obtained from eq.~(\ref{eq:L}) is positive-definite. Taking the conjugate-momentum to be $\mathfrak{p}_{z}(x)=\partial\widetilde{\mathfrak{f}}_{z}(x)/\partial t$, we find
\begin{align}
&\{\mathfrak{f}_{z}(t,\x),\widetilde{\mathfrak{f}}_{z}(t,\y)\}=O, \label{eq:fdf}\\
&\{\mathfrak{f}_{z}(t,\x),\mathfrak{f}_{z}(t,\y)\}=\{\mathfrak{p}_{z}(t,\x),\mathfrak{p}_{z}(t,\y)\}=O,\label{eq:ff}\\
&\{\mathfrak{f}_{z}(t,\x),\mathfrak{p}_{z}(t,\y)\}=i\int \frac{d^{3}p}{(2\pi)^{3}}\, e^{i\mathbf{p\cdot(x-y)}}\left[I+\mathcal{G}_{z}(\p)\right].\label{eq:fp}
\end{align}




We now explore the effects of non-locality with the $U(1)$ interaction and the fermionic self-interaction (see app.~\ref{self}). The non-local anti-commutator plays an important role which has previously not been appreciated. As for the $U(1)$ interaction, we consider the following Lagrangian in the interacting picture
\begin{equation}
\mathscr{L}=D^{\mu}\widetilde{\mathfrak{f}}D_{\mu}\mathfrak{f}-\frac{1}{4}f^{\mu\nu}f_{\mu\nu}
\end{equation}
where $a^{\mu}$ is a vector field and $f_{\mu\nu}$ the field strength tensor with 
\begin{eqnarray}
&& D_{\mu}=\partial_{\mu}-iea_{\mu},\quad D^{\mu}=\partial^{\mu}+iea^{\mu}, \\
&& f_{\mu\nu}=\partial_{\mu}a_{\nu}-\partial_{\nu}a_{\mu}.
\end{eqnarray}
The interacting Hamiltonian density is given by
\begin{equation}
\mathscr{H}=ie\left[\,\widetilde{\mathfrak{f}}(\partial^{\mu}\mathfrak{f})-(\partial^{\mu}\widetilde{\mathfrak{f}})\mathfrak{f}\right]a_{\mu}
-e^{2}(\widetilde{\mathfrak{f}}\mathfrak{f})a^{\mu}a_{\mu}.
\end{equation}
By virtue of eq.~(\ref{eq:fp}), the interacting density is in general \textit{non-local}, $[\mathscr{H}(t,\x),\mathscr{H}(t,\y)]\neq0$.	
However, one should note that the commutator  is proportional to integrals involving $a^{0}(x)a^{0}(y)$ so causality is in fact preserved in the temporal gauge where $a^{0}=0$. This feature is a consequence of Lorentz-violation. Since the objective of this paper is to explore the ramifications of Lorentz-violation, we shall explore the interacting theory in the temporal gauge where causality is preserved.

\begin{widetext}
To determine the effects of non-locality, we follow the prescription presented in app.~\ref{self} by transforming to the Heisenberg picture. In the temporal gauge, $[V(t),\mathfrak{f}(t,\x)]=0$ so we obtain eq.~(\ref{eq:Heisenberg}) and find
\begin{eqnarray}
[V(t),\partial_{t}\mathfrak{f}_{i}(t,\x)]&=&-e\int d^{3}y\sum_{j}\mathfrak{D}_{ij}(\x-\y)\left[\boldsymbol{a}(t,\y)\cdot\boldsymbol{\nabla}\mathfrak{f}_{j}(t,\y)\right]+ie\int d^{3}y\sum_{j}\mathfrak{D}_{ij}(\x-\y)\mathfrak{f}_{j}(t,\y)[\boldsymbol{a}(t,\y)\cdot\boldsymbol{a}(t,\y)]\nonumber\\
&=&-e[\boldsymbol{a}(t,\x)\cdot\boldsymbol{\nabla}\mathfrak{f}_{i}(t,\x)]+ie^{2}\mathfrak{f}_{i}(t,\x)[\boldsymbol{a}(t,\y)\cdot\boldsymbol{a}(t,\y)]
-e\int d^{3}y\sum_{j}\mathcal{G}_{ij}(\x-\y)[\boldsymbol{a}(t,\y)\cdot\boldsymbol{\nabla}\mathfrak{f}_{j}(t,\y)]\nonumber\\
&&+ie^{2}\int d^{3}y\sum_{j}\mathcal{G}_{ij}(\x-\y)\mathfrak{f}_{j}(t,\y)[\boldsymbol{a}(t,\y)\cdot\boldsymbol{a}(t,\y)]
\end{eqnarray}
where $\mathcal{G}(\x-\y)$ is the Fourier transform of $\mathcal{G}(\p)$
\begin{equation}
\mathcal{G}(\x-\y)=\frac{1}{(2\pi)^{3}}\int d^{3}p\,e^{i\mathbf{p\cdot(x-y)}}\mathcal{G}(\p).
\end{equation}
Therefore, the equation of motion for $\mathfrak{F}$ is
\begin{eqnarray}
(\partial^{\mu}\partial_{\mu}+m^{2})\mathfrak{F}_{i}(t,\x)&=&-ie\left[\boldsymbol{A}(t,\x)\cdot\boldsymbol{\nabla}\mathfrak{F}_{i}(t,\x)\right]
-e^{2}\mathfrak{F}_{i}(t,\x)[\boldsymbol{A}(t,\x)\cdot\boldsymbol{A}(t,\x)]\nonumber\\
&&-ie\int d^{3}y\sum_{j}\mathcal{G}_{ij}(\x-\y)\left[\boldsymbol{A}(t,\y)\cdot\boldsymbol{\nabla}\mathfrak{F}_{j}(t,\y)\right]
-e^{2}\int d^{3}y\sum_{j}\mathcal{G}_{ij}(\x-\y)\mathfrak{F}_{j}(t,\y)[\boldsymbol{A}(t,\y)\cdot\boldsymbol{A}(t,\y)].\nonumber\\ \label{eq:u1}
\end{eqnarray}
The interaction that appears on the right-hand side of eq.~(\ref{eq:u1}) is local in time so we may replace the fields in the Heisenberg picture by the interacting picture. We have refrained from writing down the Lagrangian for eq.~(\ref{eq:u1}) because this requires us to determine the action of the adjoint on $A^{\mu}$. In quantum electrodynamics, the gauge field and the associated Lagrangian are Hermitian. But for mass dimension one fermions, the Lagrangian satisfies the condition $\mathscr{L}=\mathscr{L}^{\ddag}$ where $\ddag$ is a non-Hermitian adjoint that acts on Elko~\cite{Lee:2019fni}. To determine $A^{\ddag}$ requires us study interactions and ensure that the scattering amplitudes satisfy certain consistency conditions so we postpone the discussions for future works. Because $\mathcal{G}(\p)\lambda^{S/A}(\p)=\pm\lambda^{S/A}(\p)$ and $\widetilde{\lambda}^{S/A}(\p)\mathcal{G}(\p)=\pm\widetilde{\lambda}^{S/A}(\p)$, when computing the scattering amplitudes, the spatial integrals reduce to momentum-conserving $\delta$-functions  so they do not directly give rise to Lorentz-violations or non-locality (see app.~\ref{self} where we explicitly computed the scattering amplitude for the self-interaction). To the best of our knowledge, the actual Lorentz-violation arises in two scenarios. One is the averaged cross-sections and decay rates where we must use the Elko spin-sums. Another scenario involves scattering amplitudes that are functions of the propagator eq.~(\ref{eq:prop}). 

Here, it is important to note that the Lorentz-violating terms are not suppressed with respect to their Lorentz-invariant counterparts. Therefore, we expect the contributions to observables from the Lorentz-invariant and violating terms to have the same order of magnitude. For cross-sections and decay rates involving mass dimension one fermions as initial states, the $\mathcal{G}$ matrix is expected to induce excess or deficit to particle emissions along some preferred directions which would give rise to interesting phenomenologies.  To derive observational signatures and put constraints on their masses and coupling constants, a new formalism of interacting mass dimension one fermions must be established. The new formalism will presented elsewhere in a separate publication.

A similar prescription also applies to gravity where the action is given by (see~\cite{daSilva:2014kfa,Pereira:2016eez,Rogerio:2019evl} for more details)
\begin{equation}
S=\int d^{4}x\sqrt{-g}(g^{\mu\nu}\nabla_{\mu}\widetilde{\mathfrak{f}}\nabla_{\nu}\mathfrak{f}-m^{2}\,\widetilde{\mathfrak{f}}\mathfrak{f}).
\end{equation}
Strictly speaking, within the present framework, non-local contributions must be included but this is not important for the discussion here. In the flat space-time limit $g_{\mu\nu}=\eta_{\mu\nu}+\kappa h_{\mu\nu}$ where $\kappa=16\pi G_{N}$, we find the commutator of the interacting density to be non-vanishing for arbitrary $h_{\mu\nu}$ at equal-time. But since the non-vanishing terms in the commutator are proportional to $h_{00}$ and $h_{0i}$, we can exploit the gauge freedom $h_{\mu\nu}\rightarrow h_{\mu\nu}+\partial_{\mu}\xi_{\nu}+\partial_{\nu}\xi_{\mu}$ by choosing $h_{00}=h_{0i}=0$ (the synchronous gauge) to preserve causality.

\end{widetext}

\section{Conclusions}

The Lounesto classification provides a new and intriguing possibility to study the physics beyond the SM. Among them, the Dirac and Weyl spinors have already found important applications in particle physics. Therefore, it is natural to ask whether the remaining spinors have any applications in particle physics. The works of~\cite{Ahluwalia:2004sz,Ahluwalia:2004ab,daRocha:2005ti} showed that a mass dimension one fermionic field naturally emerges from Elko (flag-pole spinors). Following their works, results obtained in this paper and ref.~\cite{Rogerio:2019xcu} show that mass dimension one fermions can also be constructed from flag-dipole spinors. Specifically, the flag-dipole spinors are related to Elko by a one-parameter matrix transformation $\mathcal{Z}(z)$ where $z$ is a non-zero complex number. The parameter $z$ is interpreted as an additional fermionic degeneracy and its properties require further investigation. For now, we have noted that the degeneracy can enhance spin-averaged processes and loop corrections. In other words, it can amplify the effects of Lorentz-violation so the value of $z$ should be constrained by the known dark matter phenomenologies and Lorentz symmetry.

In this paper, we focused on the non-covariant quantum field theories constructed from Elko and flag-dipole spinors. We find, the non-local anti-commutators induce non-local interactions. For the fermionic self-interaction, the $\mathcal{Z}(z)$ induces an additional fermionic degeneracy when the fermionic fields take different values of $z$.

For the $U(1)$ theory, interactions that contain time-derivative term is non-causal but this problem can be resolved by working in the temporal gauge ($a^{0}=0$). Similarly, for gravitational interactions, we can work in the synchronous gauge ($h_{00}=h_{0i}=0$) to preserve causality. In doing so we lose manifest Lorentz-invariance but this should not be too surprising. The spatially non-local interactions render the equations more complicated but as we have discussed above, they do not explicitly give rise to Lorentz-violation so it can be dealt with in a straightforward manner. Constraints on masses and coupling constants will be studied elsewhere.

\section*{Acknowledgements}

I am grateful to D.~V.~Ahluwalia for various discussions over the years that culminated the completion of this work. I am grateful to the referee for the constructive comments and criticisms. I would like to thank R. J. Bueno Rogerio for correspondence.

\appendix

\section{Fermionic degeneracy and self-interaction}\label{self}

In an interacting theory, the spin-averaged processes and loop-corrections to the propagator are dependent on the traces of $\mathcal{G}_{z}$. Since $\mathcal{G}_{z}$ is off-diagonal, the trace of their products are non-vanishing only when there are even number of them present. A direct evaluation shows that they are $z$-independent
\begin{align}
\mbox{tr}[\mathcal{G}_{z}(\p_{1})\mathcal{G}_{z}(\p_{2})\cdots\mathcal{G}_{z}(\p_{2n})]=
\mbox{tr}[\mathcal{G}(\p_{1})\mathcal{G}(\p_{2})\cdots\mathcal{G}(\p_{2n})]
\end{align}
where $\mathcal{G}(\p)=\mathcal{G}_{z}(\p)\vert_{|z|=1}$. Therefore, it would seem that mass dimension one fermions with $|z|\neq1$ and $|z|=1$ are physically equivalent.

\begin{widetext}
The above results is true if we only consider fermionic fields with a single value of $z$. A more general possibility is to study interactions between fermionic fields with different values of $z$. In this case, the traces become $z$-dependent. When $n=1,2$, we find
\begin{eqnarray}
&& \mbox{tr}[\mathcal{G}_{z_{1}}(\p_{1})\mathcal{G}_{z_{2}}(\p_{2})]=
2\left(\frac{|z_{1}|^{2}}{|z_{2}|^{2}}+\frac{|z_{2}|^{2}}{|z_{1}|^{2}}\right)\cos(\phi_{1}-\phi_{2}),\label{eq:n1}\\
&&\mbox{tr}[\mathcal{G}_{z_{1}}(\p_{1})\cdots\mathcal{G}_{z_{4}}(\p_{4})]=
2\left(\frac{|z_{1}z_{3}|^{2}}{|z_{2}z_{4}|^{2}}+\frac{|z_{2}z_{4}|^{2}}{|z_{1}z_{3}|^{2}}\right)\cos(\phi_{1}-\phi_{2}+\phi_{3}-\phi_{4}). \label{eq:n2}
\end{eqnarray}
Therefore, the magnitudes of the spin-averaged processes and loop corrections to propagators can be enhanced by choosing one of the $|z_{i}|$ to be larger than all the others. In other words, we obtain an additional fermionic degeneracy by assigning different values of $z$ to the fermionic field.
\end{widetext}

For the matrix elements of $\mathcal{Z}$ and the physical processes to be finite, $|z|$ must be non-zero and bounded. For example, for eq.~(\ref{eq:n1}), by imposing the bound
\begin{equation}
|\mbox{tr}[\mathcal{G}_{z_{1}}(\p_{1})\mathcal{G}_{z_{2}}(\p_{2})]|\leq 4|Z|^{4},\quad |Z|>1
\end{equation}
we find that $|z_{i}|$ is bounded by
\begin{equation}
|Z|^{-1}\leq |z_{i}| \leq |Z|,\quad i=1,2.
\end{equation}
Apart from these two conditions, $|z|$ can either be continuous or discrete. 
In this case, we may then consider self-interaction of the form
\begin{equation}
V_{z_{1},z_{2}}(t)=\frac{g}{2}\int d^{3}x \left\{\left[\,\widetilde{\mathfrak{f}}_{z_{1}}(x)\mathfrak{f}_{z_{2}}(x)\right]^{2}+\left[\,\widetilde{\mathfrak{f}}_{z_{2}}(x)\mathfrak{f}_{z_{1}}(x)\right]^{2}\right\}.
\end{equation}
Depending on whether $z$ is continuous or discrete, we may further generalize the interaction by summing or integrating over the index. If the index is continuous, by imposing the appropriate bounds, we have
\begin{equation}
V_{Z}(t)=\int^{|Z|}_{|Z|^{-1}}dz_{1}dz_{2} V_{z_{1}z_{2}}(t).
\end{equation}

An important issue is the effects of non-local anti-commutators on the interactions which has been neglected in the previous works. As a result, certain contributions to the interactions have been missed. To deal with this properly, we need to study the dynamics of the interacting fields. Therefore, it is necessary to transit from the free field theory to the Heisenberg picture. Here, we consider the fermionic self-interaction
\begin{equation}
V(t)=\frac{g}{2}\int d^{3}x \left[\,\widetilde{\mathfrak{f}}(x)\mathfrak{f}(x)\right]^{2}\label{eq:vf}
\end{equation}
where $\widetilde{\mathfrak{f}}$ and $\mathfrak{f}$ are free fields in the interacting picture. It is trivial to introduce $z$-dependence to eq.~(\ref{eq:vf}) if one wishes. 
Let $H=H_{0}+V$ be the full Hamiltonian with $H_{0}$ being the free part. The fermionic field in the Heisenberg picture is then given by
\begin{align}
& \mathfrak{F}(t,\x)\equiv U(t)\mathfrak{f}(t,\x)U^{-1}(t), \label{eq:fh} \\
& U(t)\equiv e^{iHt}e^{-iH_{0}t} \label{eq:ut}
\end{align} 
where the field in the interacting picture satisfies the Klein-Gordon equation
\begin{equation}
(\partial^{\mu}\partial_{\mu}+m^{2})\mathfrak{f}(x)=0.
\end{equation}
To derive the field equation for $\mathfrak{F}$, we compute its first and second time derivative. The first time derivative of $\mathfrak{F}$ is given by
\begin{eqnarray}
\partial_{t}\mathfrak{F}_{i}(t,\x)&=&U(t)\big\{\partial_{t}\mathfrak{f}_{i}(t,\x)+i[V(t),\mathfrak{f}_{i}(t,\x)]\big\}U^{-1}(t)\nonumber\\
&=&U(t)\left[\partial_{t}\mathfrak{f}_{i}(t,\x)\right]U^{-1}(t)
\end{eqnarray}
where we have used eq.~(\ref{eq:fdf}). Differentiating with respect to time again, we obtain
\begin{equation}
\partial^{2}_{t}\mathfrak{F}_{i}(t,\x)=U(t)\big\{\partial^{2}_{t}\mathfrak{f}_{i}(t,\x)+i[V(t),\partial_{t}\mathfrak{f}_{i}(t,\x)]\big\}U^{-1}(t).
\end{equation}
Using the equal-time anti-commutator
\begin{align}
&\{\partial_{t}\mathfrak{f}_{i}(t,\x),\widetilde{\mathfrak{f}}_{j}(t,\y)\}=-i\mathfrak{D}_{ij}(\x-\y),\\
&\mathfrak{D}_{ij}(\x-\y)=\frac{1}{(2\pi)^{3}}\int d^{3}p\, e^{i\mathbf{p\cdot(x-y)}}\left[\delta_{ij}+\mathcal{G}_{ij}(\p)\right]
\end{align}
and the field equation for $\mathfrak{f}$, we obtain\\
\begin{widetext}
\begin{eqnarray}
\partial^{2}_{t}\mathfrak{F}_{i}(t,\x)&=&U(t)\left\{\partial^{2}_{t}\mathfrak{f}_{i}(t,\x)-g\int d^{3}y\left[\,\widetilde{\mathfrak{f}}_{\ell}(t,\y)\mathfrak{f}_{\ell}(t,\y)\right]\left[\mathfrak{D}_{ij}(\x-\y)\mathfrak{f}_{j}(t,\y)\right]\right\}U^{-1}(t) \nonumber\\
&=&U(t)\left\{-\partial_{i}\partial^{i}\mathfrak{f}_{i}(t,\x)-m^{2}\mathfrak{f}_{i}(t,\x)-g\int d^{3}y\, \left[\,\widetilde{\mathfrak{f}}_{\ell}(t,\y)\mathfrak{f}_{\ell}(t,\y)\right]\left[\mathfrak{D}_{ij}(\x-\y)\mathfrak{f}_{j}(t,\y)\right]\right\}U^{-1}(t)\nonumber\\
&=&\left\{-\partial_{i}\partial^{i}\mathfrak{F}_{i}(t,\x)-m^{2}\mathfrak{F}_{i}(t,\x)-g\int d^{3}y\, \left[\,\widetilde{\mathfrak{F}}_{\ell}(t,\y)\mathfrak{F}_{\ell}(t,\y)\right]\left[\mathfrak{D}_{ij}(\x-\y)\mathfrak{F}_{j}(t,\y)\right]\right\}
\end{eqnarray}
Therefore, the field equation for $\mathfrak{F}$ is given by
\begin{eqnarray}
(\partial^{\mu}\partial_{\mu}+m^{2})\mathfrak{F}_{i}(t,\x)&=&-g\int d^{3}y\left[\widetilde{\mathfrak{F}}_{\ell}(t,\y)\mathfrak{F}_{\ell}(t,\y)\right]\left[\mathfrak{D}_{ij}(\x-\y)\mathfrak{F}_{j}(t,\y)\right]. \label{eq:Heisenberg}
\end{eqnarray}
Similarly, the field equation for $\widetilde{\mathfrak{F}}(t,\x)$ is
\begin{eqnarray}
(\partial^{\mu}\partial_{\mu}+m^{2})\widetilde{\mathfrak{F}}_{i}(t,\x)&=&-g\int d^{3}y\left[\widetilde{\mathfrak{F}}_{\ell}(t,\y)\mathfrak{F}_{\ell}(t,\y)\right]\left[\widetilde{\mathfrak{F}}_{j}(t,\y)\mathfrak{D}_{ji}(\x-\y)\right].
\end{eqnarray}
Therefore, the Lagrangian is given by
\begin{eqnarray}
\mathfrak{L}(t,\x)&=&\partial^{\mu}\widetilde{\mathfrak{F}}(t,\x)\partial_{\mu}\mathfrak{F}(t,\x)-m^{2}\widetilde{\mathfrak{F}}(t,\x)\mathfrak{F}(t,\x)-\frac{g}{2}\left[\widetilde{\mathfrak{F}}(t,\x)\mathfrak{F}(t,\x)\right]^{2}\nonumber\\
&&-g\int d^{3}y\left[\widetilde{\mathfrak{F}}(t,\y)\mathfrak{F}(t,\y)\right]\left[\widetilde{\mathfrak{F}}(t,\x)\mathcal{G}(\x-\y)\mathfrak{F}(t,\y)+\widetilde{\mathfrak{F}}(t,\y)\mathcal{G}(\x-\y)\mathfrak{F}(t,\x)\right]
\end{eqnarray}
There is an important point to note here. If we compare eq.~(\ref{eq:Heisenberg}) to (\ref{eq:ut}), we find that the non-local interaction that contributes to the $S$-matrix is absent in $U$. This is not a contradiction but a subtlety that arises from the non-local anti-commutators which has been neglected in the previous works. In fact, from the above derivation, we see that unless the field and its conjugate variables satisfy the canonical anti-commutators, the Hamiltonian that appears in $U$ will not coincide with its counterpart in the field equation. 
The integral terms that appear in the interaction are non-local in space but local in time. Therefore, we may transit from the Heisenberg to the interacting picture by replacing $\mathfrak{F}$ with $\mathfrak{f}$. Because the self-interaction are functions of $\widetilde{\mathfrak{f}}\mathfrak{f}$, the interacting density is local so the causality of the $S$-matrix is preserved.

As an example, we compute the $S$-matrix for the process $12\rightarrow34$ at tree-level. 
For this purpose, let us write the $S$-matrix as
\begin{equation}
S_{(34)(12)}=S^{(1)}_{(34)(12)}+S^{(2)}_{(34)(12)}
\end{equation}
where $S^{(1)}_{(34)(12)}$ and $S^{(2)}_{(34)(12)}$ represent the contributions from the local and non-local interactions respectively. The local contribution evaluates to
\begin{equation}
S^{(1)}_{(34)(12)}=-\frac{ig\delta^{4}(p_{3}+p_{4}-p_{1}-p_{2})}{(2\pi)^{2}\sqrt{16m^{4}E_{1}E_{2}E_{3} E_{4}}}
\left[(\widetilde{\lambda}^{S}_{4}\lambda^{S}_{1})(\widetilde{\lambda}^{S}_{3}\lambda^{S}_{2})-(\widetilde{\lambda}^{S}_{4}\lambda^{S}_{2})(\widetilde{\lambda}^{S}_{3}\lambda^{S}_{1})\right]\label{eq:s1}
\end{equation}
where $\lambda^{S}_{i}=\lambda^{S}(\p_{i},\sigma_{i})$. A similar calculation for the non-local contribution yields
\begin{eqnarray}
S^{(2)}_{(34)(12)}&=&-\frac{ig\delta^{4}(p_{3}+p_{4}-p_{1}-p_{2})}{(2\pi)^{2}\sqrt{16m^{4}E_{1}E_{2}E_{3} E_{4}}}
{\Big\{}\left[(\widetilde{\lambda}^{S}_{4}\lambda^{S}_{1})(\widetilde{\lambda}^{S}_{3}\mathcal{G}_{3}\lambda^{S}_{2})+(\widetilde{\lambda}^{S}_{1}\lambda^{S}_{4})(\widetilde{\lambda}^{S}_{2}\mathcal{G}_{3}\lambda^{S}_{3})\right]-\left[(\widetilde{\lambda}^{S}_{4}\lambda^{S}_{2})(\widetilde{\lambda}^{S}_{3}\mathcal{G}_{3}\lambda^{S}_{1})+(\widetilde{\lambda}^{S}_{2}\lambda^{S}_{4})(\widetilde{\lambda}^{S}_{1}\mathcal{G}_{3}\lambda^{S}_{3})\right]\nonumber\\
&&-\left[(\widetilde{\lambda}^{S}_{3}\lambda^{S}_{1})(\widetilde{\lambda}^{S}_{4}\mathcal{G}_{4}\lambda^{S}_{2})+(\widetilde{\lambda}^{S}_{1}\lambda^{S}_{3})(\widetilde{\lambda}^{S}_{2}\mathcal{G}_{4}\lambda^{S}_{4})\right]
+\left[(\widetilde{\lambda}^{S}_{3}\lambda^{S}_{2})(\widetilde{\lambda}^{S}_{4}\mathcal{G}_{4}\lambda^{S}_{1})+(\widetilde{\lambda}^{S}_{2}\lambda^{S}_{3})(\widetilde{\lambda}^{S}_{1}\mathcal{G}_{4}\lambda^{S}_{4})\right]
{\Big\}}\label{eq:s2}
\end{eqnarray}
where $\mathcal{G}_{i}=\mathcal{G}(\p_{i})$. Using the identities $\mathcal{G}_{i}\lambda^{S}_{i}=\lambda^{S}_{i}$ and $\widetilde{\lambda}^{S}_{i}\mathcal{G}_{i}=\widetilde{\lambda}^{S}_{i}$, equation~(\ref{eq:s2}) simplifies to
\begin{equation}
S^{(2)}_{(34)(12)}=-\frac{2ig\delta^{4}(p_{3}+p_{4}-p_{1}-p_{2})}{(2\pi)^{2}\sqrt{16m^{4}E_{1}E_{2}E_{3} E_{4}}}
\left\{\left[(\widetilde{\lambda}^{S}_{4}\lambda^{S}_{1})(\widetilde{\lambda}^{S}_{3}\lambda^{S}_{2})
+(\widetilde{\lambda}^{S}_{1}\lambda^{S}_{4})(\widetilde{\lambda}^{S}_{2}\lambda^{S}_{3})\right]
-\left[(\widetilde{\lambda}^{S}_{4}\lambda^{S}_{2})(\widetilde{\lambda}^{S}_{3}\lambda^{S}_{1})
+(\widetilde{\lambda}^{S}_{2}\lambda^{S}_{4})(\widetilde{\lambda}^{S}_{1}\lambda^{S}_{3})\right]\right\}.\label{eq:ss2}
\end{equation}
Comparing eqs.~(\ref{eq:s1}) and (\ref{eq:ss2}), we see that apart from the different normalization factors, $S^{(1)}_{(34)(12)}$ and $S^{(2)}_{(34)(12)}$ are of the same form. In this sense, the interactions involving $\mathcal{G}(\x-\y)$ are local for the process $12\rightarrow34$ at tree-level. The cross-sections and loop corrections will be considered in a separate publication.


\end{widetext}

%
\bibliography{Bibliography}
\bibliographystyle{unsrt}

%

\end{document}